\documentclass[12pt,superscriptaddress,aps,prd,preprint]{revtex4}

\usepackage{amsfonts}
\usepackage{slashed}
\usepackage{amsmath}
\newcommand{\bea}{\begin{eqnarray}}
\newcommand{\eea}{\end{eqnarray}}

\newcommand{\pa}{\partial}

\title{On plane wave solutions in Lorentz-violating extensions of gravity}

\begin{document}
	
	\title{On plane wave solutions in Lorentz-violating extensions of gravity}

	\author{J. R. Nascimento}
	\author{A. Yu. Petrov}
	\affiliation{Departamento de F\'{\i}sica, Universidade Federal da Para\'\i ba,\\
		Caixa Postal 5008, 58051-970, Jo\~ao Pessoa, Para\'\i ba, Brazil}
	\email{jroberto, petrov@fisica.ufpb.br}
	\author{A. R. Vieira}
	\affiliation{Universidade Federal do Tri\^{a}ngulo Mineiro,\\ 
		Campus Iturama, 38280-000, Iturama, MG, Brazil}
	\email{alexandre.vieira@uftm.edu.br}
	
	
	
	
	
	
	
	
	\begin{abstract}
		In this paper, we obtain dispersion relations corresponding to plane wave solutions in Lorentz-breaking extensions of gravity 
		with dimension 3, 4, 5 and 6 operators. We demonstrate that these dispersion relations display a usual Lorentz-invariant mode when the 
		corresponding additive term involves higher derivatives.
	\end{abstract}

\maketitle

\section{Introduction}

The observations of gravitational waves performed within the LIGO experiment \cite{LIGO} certainly represent themselves as one of the most 
important experimental 
confirmations of general relativity. At the same time, various modifications and extensions of gravity are being discussed now. The main 
motivations for these extensions are, first, need in development of a perturbatively consistent gravity model, which is expected to be both 
renormalizable and ghost-free, second, necessity to explain the cosmic acceleration originally reported in \cite{Riess}. At the same time, the 
concept of Lorentz symmetry breaking, possessing various motivations -- string theory, minimal length, quantum fluctuations of geometry, loop 
quantum gravity etc. -- clearly can be implemented within the 
gravity context, and the Lorentz-violating (LV) standard model extension (LV SME) \cite{coll1,coll2} was generalized to include gravity in \cite
{KosGra}. All these studies clearly establish the question about possible gravitational wave solutions in LV extended gravity models. It is well 
known that in Lorentz-breaking extensions of other theories, e.g. the electrodynamics, plane wave solutions display nontrivial behavior, such as 
birefringence and rotation of the polarization plane in the vacuum (see e.g. \cite{coll1,coll2}), therefore, it is natural to search for such 
phenomena also in the gravitational wave case.

First study of the plane wave solutions in LV gravity has been performed in \cite{JaPi} where the four-dimensional Chern-Simons (CS) modified 
gravity presenting the Lorentz-breaking behavior for the special form of the CS coefficient $\vartheta=k_{\mu}x^{\mu}$ was considered. However, 
it turns to be that the only consistent plane wave solution in this theory displays only usual, Lorentz-invariant dispersion relations, with the 
intensities of two polarizations for gravitational waves are different. A more interesting situation takes place in \cite{ourgra} where the 
additive one-derivative LV term breaks the gauge symmetry (for a detailed discussion of gauge symmetry breaking in gravity see \cite{KosGB}) -- 
in this case two polarizations propagate with distinct phase velocities depending on the Lorentz-breaking parameter and different from the speed 
of light.

Therefore, the natural problem consists in studying of plane wave solutions in gravity theories with various recently proposed LV additive terms 
\cite{KosLi,Bailey}. This issue will be discussed in the present paper.

The structure of the paper looks like follows. In section 2, we consider the dispersion relations in modified gravity models representing 
themselves as a sum of the usual Einstein-Hilbert action and terms introduced in \cite{KosLi}. In section 3, we obtain the dispersion 
relations in theories whose action is given by a sum of the Einstein-Hilbert term and some new linearized gauge invariant terms. Finally, in 
section 4 we summarize our results.
 
\section{Dispersion relations for full-fledged LV terms in gravitational sector}

In this section we consider the dispersion relations generated by additive full-fledged LV terms in gravitational sector, proposed in \cite{KosLi}.
Our starting point is the following decomposition of the $h_{\mu\nu}$ tensor into its irreducible components originally introduced in \cite{JaPi} 
(see also \cite{ourgra}):
\bea
\label{decomp}
h^{00}=n, \quad\, h^{0i}=n^i_T+\pa^in_L;\nonumber\\
h^{ij}=(\delta^{ij}-\frac{\pa^i\pa^j}{\nabla^2})\phi+\frac{\pa^i\pa^j}{\nabla^2}\chi+(\pa^i\xi^j_T+\pa^j\xi^i_T)+h^{ij}_{TT},
\label{eqh}
\eea
where $h^{ij}_{TT}\equiv \tilde{h}_{ij}$ is transverse and traceless and $\xi^i_T$ is transverse. Our signature is $(+,-,-,-)$. In this case, the 
Lagrangian for a spin-2 field of the linearized gravity (see e. g. \cite{Veltman})
\bea
\label{FP}
{\cal L}_0=\frac{1}{4}\partial_{\mu}h_{\alpha}^{\alpha}\partial^{\mu}h_{\beta}^{\beta}-\frac{1}{2}\partial_{\beta}h_{\alpha}^{\alpha}\partial^{\mu
}h^{\beta}_{\mu}-\frac{1}{4}\partial_{\mu}h_{\alpha\beta}\partial^{\mu}h^{\alpha\beta}+\frac{1}{2}\partial_{\alpha}h_{\nu\beta}\partial^{\nu}h^{
\alpha\beta},
\eea
which is nothing more than the well known Einstein-Hilbert Lagrangian for the weak field, takes the form (see e.g. \cite{JaPi}):
\bea
\label{fpfis}
{\cal L}_{FP}=-\frac{1}{4}\tilde{h}^{ij}\Box \tilde{h}^{ij}+\frac{1}{2}\phi\Box\phi+\frac{1}{2}(\pa^i\sigma^j_T)^2+\phi\Lambda,
\eea
where $\sigma^i_T=n^i_T+\dot{\xi}^i_T$ is transverse and $\Lambda=\nabla^2(n+2\dot{n}_L)+\ddot{\chi}$ is the lagrange multiplier. These quantities
show that $\tilde{h}_{ij}$ is the only propagating field \cite{JaPi}.

For this Lagrangian, the dispersion relations for the only physical modes presented by $\tilde{h}_{ij}$, are the usual ones, $E^2=\vec{p}^2$, 
as it must be.
 
So, let us perform the similar decomposition for various additive LV terms introduced in the table VI given in \cite{KosLi} in the linearized 
case, with dimensions of these terms up to 6. Since we are interested in dynamics of $\tilde{h}_{ij}$ which is traceless, we can assume 
$\sqrt{|g|}=1$. Also, in the linearized case we can require the external vectors (tensors) 
$(\stackrel{\smile}{k}^{(n)})^{\mu_1\ldots\mu_{n-2}}\equiv (k^{(n)})^{\mu_1\ldots\mu_{n-2}}$ to be (approximately) constant in order to avoid 
non-constant free parameters in dispersion relations (i.e. to require that only gravitational fields can propagate), so, all derivatives of 
external vectors (tensors) are disregarded. From the physical viewpoint, this condition is consistent with the conservation of the 
energy-momentum tensor since it corresponds to homogeneity of the space-time.

From now, our methodology is as follows. For any additive Lorentz-breaking term, we will keep only its observable (transverse-traceless) components, obtaining thus extra contributions to Lagrangians of these components, and study the propagation of the plane waves described by these physical components, and corresponding dispersion relations. For the sake of simplicity, we assume that the Lorentz-breaking tensor parameters can be completely characterized by one Lorentz-breaking vector (or pseudovector), similarly to aether terms \cite{Carroll,aether}.
 
The simplest example of the LV parameter in gravity given in \cite{KosLi} is  $(\stackrel{\smile}{k}_{\Gamma}^{(3)})^{\mu}\equiv (k^{(3)})^{\mu}$ 
defining the dimension-3 operator 
$(k^{(3)})^{\mu}\Gamma^{\alpha}_{\mu\alpha}$. In the linearized case, we can write
 $$
 \Gamma^{\alpha}_{\mu\alpha}=-\frac{1}{2}h^{\alpha\gamma}\partial_{\mu}h_{\alpha\gamma}+O(h^3).
 $$

Let us assume that our plane gravitational wave propagates along $x_3=z$ axis, i.e. $h_{\mu\nu}=\tilde{h}_{\mu\nu}e^{i(Et-pz)}$. In this case, 
there will be 
no second derivatives acting on any components of decomposition of $h_{\mu\nu}$ except of the usual transverse-traceless $\tilde{h}_{ij}$. 
Similarly to \cite
{JaPi,ourgra}, we can define two polarizations of $\tilde{h}_{ij}$ as follows: $\tilde{h}_{11}=-\tilde{h}_{22}=T$, $h_{12}=h_{21}=S$, all other 
components of $\tilde{h}_{ij}$ are 
zero.

First of all, in this case we have (with $(k^{(3)})^{\mu}\equiv k^{\mu}$)
 \bea
 \label{l3}
 {\cal L}^{(3)}=-\frac{1}{2}(k^{(3)})^{\mu}\tilde{h}^{ij}\partial_{\mu}\tilde{h}_{ij}=-\frac{1}{2}(k^0\tilde{h}^{ij}\partial_0\tilde{h}_{ij}+k^3
\tilde{h}^{ij}
\partial_3\tilde{h}_{ij}),
 \eea
where we disregarded all other components of $h_{\mu\nu}$. We immediately see that this term is evidently a total derivative, hence its impact to 
the modified linearized equations of motion is trivial, thus, adding of ${\cal L}^{(3)}$ will not affect plane wave solutions independently of 
the direction of the vector $(k^{(3)})^{\mu}$. Unlike (\ref{l3}), the term
$\epsilon^{\mu\nu\lambda\rho}b_{\mu}h_{\nu\alpha}\partial_{\lambda}h_{\rho}^{\alpha}$ discussed in \cite{ourgra}, being also dimension-3 term, is 
described by a pseudo-vector $b_{\mu}$, and in this case the dispersion relations are different, so that for $b^{\mu}=-\frac{b}{2}\hat{z}$, i.e. 
the LV vector is parallel to the wave direction, they look like $(E\pm b)^2-(p+b)^2=0$ which implies the group velocity less than the speed of 
light \cite{ourgra}. Moreover, this term cannot be expressed in terms of usual geometric objects, such as a connection or a curvature, hence it is 
apparently well defined only within a linearized gravity but not in a full-fledged one. We note that this term breaks the gauge symmetry,  and this fact establishes the natural question about impact of breaking the gauge symmetry on the dispersion relations (see \cite{KosGB} for a discussion of violating the general covariance in gravity). In this section and in the 
next one, we will work both in gauge-breaking and gauge-invariant scenarios in order to see if the breaking of gauge invariance implies unusual dispersion relations.

For studying of higher-order terms, it is useful to write down lower-order contributions to Riemann and Ricci tensors. For the Riemann tensor we have (cf. \cite
{Veltman})
\bea
R_{\mu\nu\alpha\beta}&=&\frac{1}{2}(\partial_{\nu}\partial_{\alpha}h_{\mu\beta}-\partial_{\mu}\partial_{\alpha}h_{\nu\beta}-\partial_{\nu}
\partial_{\beta}h_{
\mu\alpha}+\partial_{\mu}\partial_{\beta}h_{\nu\alpha})+\nonumber\\
&+&\partial_{\alpha}\Gamma^{(2)}_{\mu,\nu\beta}-\partial_{\beta}\Gamma^{(2)}_{\mu,\nu\alpha}+
\Gamma^{(1)\gamma}_{\beta\nu}\Gamma^{(1)}_{\mu,\gamma\alpha}-
\Gamma^{(1)\gamma}_{\alpha\nu}\Gamma^{(1)}_{\mu,\gamma\beta},
\eea
where $\Gamma^{(1,2)\gamma}_{\beta\nu}$ are first and second orders in expansions of Christoffel symbols in $h_{\alpha\beta}$, explicitly,
\bea
\Gamma^{(1)\alpha}_{\beta\gamma}&=&\frac{1}{2}(\partial_{\beta}h^{\alpha}_{\gamma}+\partial_{\gamma}h^{\alpha}_{\beta}-\partial^{\alpha}h_{
\beta\gamma});\nonumber\\
\Gamma^{(2)\alpha}_{\beta\gamma}&=&-\frac{1}{2}h^{\alpha\delta}(\partial_{\beta}h_{\gamma\delta}+\partial_{\gamma}h_{\beta\delta}-\partial_{\delta
}h_{\beta\gamma}).
\eea
Our next example is the dimension-4 term $(\stackrel{\smile}{k}_{R}^{(4)})^{\mu\nu\rho\sigma}R_{\mu\nu\rho\sigma}$. The importance of this term 
consists in 
the fact that this is the simplest CPT-even LV term in gravity which for a special "aether-like" form of $(\stackrel{\smile}{k}_{R}^{(4)})^{\mu\nu\rho\sigma}$ given by 
$(\stackrel{\smile}{k}_{R}^{(4)})^{\mu\nu\rho\sigma}=u^{\mu}u^{\rho}\eta^{\nu\sigma}-u^{\mu}u^{\sigma}\eta^{\nu\rho}+u^{\nu}u^{\sigma}\eta^{\mu\rho
}-u^{\nu}u^{
\rho}\eta^{\mu\sigma}$ is reduced to the gravitational aether term introduced in \cite{Carroll}. Some studies of dispersion relations in this 
theory have 
been performed in \cite{MalufGrav}, where causality and unitarity are analyzed within the context of the bumblebee gravity for space-like and 
time-like backgrounds of the bumblebee field. Explicitly, it is demonstrated that there are two graviton dispersion relations, $p^2+\xi(b\cdot 
p)^2=0$ and $(b\cdot p)^2-b^2p^2=0$, where $b_{\mu}$ is the LV constant vector (actually, it is the v.e.v. of the bumblebee field), and $\xi$ is 
the known bumblebee-gravity coupling (see \cite{KosGra}). We note that the first dispersion relation is rather standard one for massless CPT-even 
LV theories, it arises for example in aether-like CPT-even models of scalar and gauge fields \cite{Carroll,aether}. As for the second relation, 
it corresponds to breaking the unitarity, and the energy strongly depends on direction of propagation \cite{MalufGrav}. For the vector field, 
such a relation has been obtained in \cite{Guima} for a non-canonical gauge theory where the aether term is not suppressed in comparison with the 
Maxwell term. It is natural to expect that for a generic form of $(\stackrel{\smile}{k}_{R}^{(4)})^{\mu\nu\rho\sigma}R_{\mu\nu\rho\sigma}$ 
dispersion relations do not differ essentially.

So, it is especially interesting to study the higher-derivative LV extensions of gravity. Higher-derivatives reveal informations about the whole
theory and are one of the ways to attain renormalizability \cite{Stelle}. It is well known \cite{Stelle} that in higher-derivative 
Lorentz-invariant theories ghost states arise. Their presence makes the theory unstable. However, Lorentz-breaking 
higher-derivative terms in certain cases, for example when higher derivatives are purely spatial one, do not display ghost states \cite{Myers}. Therefore, the cases where the higher time derivatives are 
ruled out due to the appropriate choice of Lorentz-breaking parameters are certainly of special interest. Also, four derivative terms are 
considered in cosmological models to explain cosmic acceleration \cite{Starobinsky}. The most known example of the dimension-5 terms is the 
gravitational CS term whose dispersion relations have been discussed in \cite{JaPi} and proved to be usual ones $E^2=\vec{p}^2$, although the CPT 
breaking manifested itself through difference of intensities for two circular polarizations (various issues related to the linearized 
gravitational CS term are also discussed in \cite{CSlin,Felipe,Brett}). One more CS-like term from the table VI given in \cite{KosLi}, is 
proportional to two Levi-Civita symbols, but it vanishes within the metric formalism since contractions like $\epsilon_{abcd}\omega_{\mu}^{ab}$ 
which are present within this term, for a Riemannian connection are equal to zero.

It remains to study the dimension-5 term proportional to $D_{\kappa}R_{\rho\sigma\mu\nu}$.
There is a number of ways to decompose tensor $(\stackrel{\smile}{k}_{D}^{(5)})^{\rho\sigma\mu\nu\kappa}$. If it is completely symmetric, this 
term is evidently ruled out due to antisymmetry of the Riemann tensor with respect to some indices, thus, the equations
of motion are reduced to the Einstein ones, hence the dispersion relations again have the usual form $E^2=\vec{p}^2$. To obtain a nontrivial 
impact of this term within the dispersion relations context, we can
decompose the dimension-5 coefficient as 
$(\stackrel{\smile}{k}_{D}^{(5)})^{\rho\sigma\mu\nu\kappa}=k^{\rho}(k^{\sigma}k^{\nu}\eta^{\mu\kappa}-k^{\sigma}k^{\kappa}\eta^{\mu\nu}+k^{\kappa}k
^{\mu}\eta^{\sigma\nu}-k^{\mu}k^{\nu}\eta^{\sigma\kappa})$, so, it has an aether-like structure being completely characterized by one vector. After  making the contraction, we find the following additional term is the linearized Einstein equations:
\begin{eqnarray}
&G^{(5)}_{\alpha\beta}=\frac{1}{4}\Big(\frac{1}{2}\eta_{\alpha\beta}k^{\kappa}k^{\rho}(k\cdot\partial )\Box h_{\rho\kappa}+k_{\alpha}k_{\beta}
\partial^{\kappa}
\partial^{\rho}(k\cdot \partial ) h_{\kappa\rho}+k^{\mu} \partial_{\alpha}(k \cdot \partial )^2 h_{\beta\mu}- \nonumber\\
&-k^{\kappa}k^{\rho} \partial_{\beta}\partial_{\alpha}(k \cdot \partial ) h_{\rho\kappa}-k_{\beta}\partial^{\rho}(k \cdot \partial )^2 h_{
\alpha\rho}+(\alpha 
\leftrightarrow \beta) \Big).
\label{eqG}
\end{eqnarray}

Considering the decomposition of the metric perturbation in eqs. (\ref{eqh}) and replacing it in eq. (\ref{eqG}), we find the additional 
terms in the equation of motion derived from Lagrangian in eq. (\ref{FP}), which, in the sector of the physical components $\tilde{h}_{ij}$, 
assumes the form
\begin{eqnarray}
&\Box \tilde{h}_{ij}+\frac{1}{2}\eta_{ij}k^a k^b (k\cdot \partial )\Box \tilde{h}_{ab}+ k_{i}k_{j}(k\cdot \partial )\partial^a \partial^b 
\tilde{h}_{ab}+\nonumber\\
&+\frac{1}{2}\partial_{i}(k\cdot \partial )^2 k^a \tilde{h}_{j a}+\frac{1}{2}\partial_{j}(k\cdot \partial )^2 k^a \tilde{h}_{i a}-k^ak^b\partial_ i
\partial_ j(k\cdot \partial )\tilde{h}_{ab}-\nonumber\\
&-\frac{1}{2}k_j\partial^a (k\cdot \partial )^2\tilde{h}_{ia}-\frac{1}{2}k_i\partial^a (k\cdot \partial )^2\tilde{h}_{ja}+(\ldots)=0.
\label{eqh2}
\end{eqnarray}
Here, dots are for the terms which do not depend on $\tilde{h}_{ik}$. For the further study, it is important to note that, first, all such terms are accompanied by Lorentz-breaking constant vectors (tensors) known to be small, can be treated effectively as small sources in corresponding wave equations for $\tilde{h}_{ij}$, and thus affect only higher-order contributions to the plane wave solutions, second, do not influence on the equations and dispersion relations for relevant, transverse-traceless components of $h_{ij}$.
Again, we consider the plane wave solutions, $h_{ij}=\tilde{h}_{ij}e^{ipx}$. As we already have done throughout this text, let us now disregard the terms proportional to $\partial^a\tilde{h}_{ab}$ and its derivatives which vanish in our case. We have
	\begin{eqnarray}
		&\Box \tilde{h}_{ij}+\frac{1}{2}\eta_{ij}k^a k^b (k\cdot \partial )\Box \tilde{h}_{ab}+\nonumber\\
		&+\frac{1}{2}\partial_{i}(k\cdot \partial )^2 k^a \tilde{h}_{j a}+\frac{1}{2}\partial_{j}(k\cdot \partial )^2 k^a \tilde{h}_{i a}-k^ak^b
\partial_ i	\partial_ j(k\cdot \partial )\tilde{h}_{ab}+(\ldots)=0.
		\label{eqh2a}
	\end{eqnarray}
 Here, as well as in next equations, the dots are for contributions to the effective equations of motion which do not depend on $\tilde{h}_{ij}$ and hence do not affect the dispersion relations.
Just as within considering the dimension-3 
term (see the discussion above and in \cite{ourgra}), we can assume $\tilde{h}_{ij}$ to have two 
polarizations states, given by $\tilde{h}_{11}=-\tilde{h}_{22}=T$ and $\tilde{h}_{12}=\tilde{h}_{21}=S$. In this case, the dispersion relations are again the usual ones $E^2=\vec{p}^2$,  for $\vec{k}$ either parallel or 
orthogonal to the wave vector. The same conclusion is valid for a generic direction of $\vec{k}$ since our plane wave depends on $t$ and $z=x_3$ only, and the terms in the second line of the equation above will not modify the dispersion relations for physical components $h_{11,12,22}$, here we remind that all other components of $h_{\mu\nu}$ do not describe physical degrees of freedom and hence can be put to zero. We conclude that presence of higher-derivative LV terms implies arising of the unique dispersion relation $E^2=\vec{p}^2$.

We can continue with studying remaining terms from table VI in \cite{KosLi}. The next operator to study is the dimension-6 one 
$D_{\kappa}D_{\lambda}R_{\rho\sigma\mu\nu}$, i.e. there is one
more partial derivative comparing with the previous term, we can use the relation $\delta(D_{\kappa}D_{\lambda}R^{\rho}_{\ \sigma\mu\nu})= \delta 
\Gamma^{\rho}_{\kappa\tau}\partial_{\lambda}R^{\tau}_{\ \sigma\mu\nu}+\Gamma^{\rho}_{\kappa\tau}\partial_{\lambda}\delta R^{\tau}_{\ 
\sigma\mu\nu}+O(h^3)$.  Similarly to the above calculations, we can also decompose the dimension-6 coefficient in the aether-like form \\
$(\stackrel{\smile}{k}^{(6)}_{D})^{\rho\sigma\mu\nu\kappa\lambda}=k^{\mu}k^{\lambda}(k^{\rho}k^{\kappa}\eta^{\nu\sigma}-
k^{\sigma}k^{\alpha}\eta^{\kappa\rho}+k^{\kappa}k^{\sigma}\eta^{\nu\rho}-k^{\kappa}k^{\rho}\eta^{\nu\sigma})$. In this case, as we would expect,
the equation of motion contains an additive term involving one more derivative and one more degree of the momentum. Explicitly, this fourth-derivative term looks like:
\begin{eqnarray}
&G^{(6)}_{\alpha\beta}=\frac{1}{4}\Big(\frac{1}{2}\eta_{\alpha\beta}k^{\kappa}k^{\rho}(k\cdot\partial )^2\Box h_{\rho\kappa}+k_{\alpha}k_{\beta}
\partial^{\kappa}\partial^{\rho}(k\cdot \partial )^2 h_{\kappa\rho}+k^{\mu} \partial_{\alpha}(k \cdot \partial )^3 h_{\beta\mu}- \nonumber\\
&-k^{\kappa}k^{\rho} \partial_{\beta}\partial_{\alpha}(k \cdot \partial )^2 h_{\rho\kappa}-k_{\beta}\partial^{\rho}(k \cdot \partial )^3 h_{
\alpha\rho}+( \alpha \leftrightarrow \beta) \Big).
\label{eqG2}
\end{eqnarray}

Eq. (\ref{eqG2}) demonstrates arising of additional terms in the equation of motion derived from Lagrangian in eq. (\ref{FP}), and in the sector 
of the physical components it takes the form:
\begin{eqnarray}
&\Box \tilde{h}_{ij}+\frac{1}{2}\eta_{ij}k^ak^b(k\cdot \partial )^2\Box \tilde{h}_{ab}+\nonumber\\
&+\frac{1}{2}k^a \partial_i (k \cdot \partial )^3\tilde{h}_{ja}- k^a k^b \partial_i \partial_j (k\cdot \partial )^2\tilde{h}_{ab}
+\frac{1}{2}k^a \partial _j(k\cdot \partial )^3\tilde{h}_{ia}+(\ldots)=0.
\end{eqnarray}
In the same way as above, we substitute plane wave solutions in the above equation. As in the dimension-5 term, if $\tilde{h}_{ij}$ has two 
polarization states, we do not find any additional term in the dispersion relation and again have $E^2=\vec {p}^2$. So, either for $\vec{k}$ 
parallel or orthogonal to the wave vector $\vec{p}$ we arrive at the usual dispersion relation. 

It remains to study the last dimension-6 term $(\stackrel{\smile}{k}_{R}^{(6)})^{
\alpha\beta\gamma\delta\mu\nu\zeta\lambda} R_{\alpha\beta\gamma\delta} R_{\mu\nu\zeta\lambda}$ from table VI of \cite{KosLi}. It is also possible 
to decompose this coefficient in the aether-like form
$(\stackrel{\smile}{k}_{R}^{(6)})^{\alpha\beta\gamma\delta\mu\nu\zeta\lambda}=k^{\alpha}k^{\beta}k^{\gamma}(k^{\lambda}k^{\nu}k^{\sigma}\eta^{\zeta
\mu}-k^{\lambda}k^{\mu}k^{\sigma}\eta^{\zeta\nu}+k^{\zeta}k^{\mu}k^{\sigma}\eta^{\lambda\nu}-k^{\zeta}k^{\nu}k^{\sigma}\eta^{\lambda\mu})$. This
term leads to the following additive term to the modified Einstein tensor:
\begin{eqnarray}
&G^{(6)}_{\alpha\beta}=(k\cdot \partial)^3 k^{\lambda}k^2(\partial_{\alpha}h_{\lambda\beta})-(k\cdot \partial )^2 k^{\lambda}k^{\zeta}k^2
\partial_{\alpha}\partial_{\beta}h_{\lambda\zeta}-(k\cdot \partial )^2 k^{\lambda}k_{\alpha}k^2\Box h_{\lambda\beta}+\nonumber\\
&+(k\cdot \partial )k^2 k^{\lambda}k^{\zeta}k_{\alpha}\partial_{\beta}\Box h_{\lambda\zeta}+(\alpha \leftrightarrow \beta)
\label{eqG4}
\end{eqnarray}

Now the equation of motion is given by
\begin{eqnarray}
&\Box \tilde{h}_{ij}-2(k \cdot \partial)^2 k^2 k^l k_i \Box \tilde{h}_{lj}+2(k\cdot \partial)k^2 k^l k^m k_{i}\partial_{j} \Box \tilde{h}_{lm}-
\nonumber\\
&-2(k \cdot \partial)^2 k^2 k^l k^m \partial_j \partial_i \tilde{h}_{lm}+2(k\cdot \partial)^3 k^2 k^l \partial_i \tilde{h}_{lj}+(i \leftrightarrow 
j)+(\ldots)=0
\end{eqnarray}

In this case, if $k$ is a space-like vector, parallel or orthogonal the wave vector, the dispersion relations are again the usual ones 
$E^2=\vec{p}^2$. 

\section{Dispersion relations for linearized gauge invariant LV terms}

	Now, let us present another approach to study of dispersion relations in linearized gravity. In this case we start with the quadratic action 
instead of the full-fledged one, but assume its invariance under the gauge transformations of the metric fluctuation $h_{\mu\nu}$ of the usual form
	\bea
	\label{gatra}
	\delta h_{\mu\nu}=\pa_{\mu}\xi_{\nu}+\pa_{\nu}\xi_{\mu},
	\eea
	with $\xi_m$ is a parameter of transformations. 

So, let us find fourth-order linearized gauge invariant terms which at the same time, being constructed on the base of the Einstein tensors in order to guarantee the gauge invariance, can be expressed in terms of the Ricci tensor and the scalar curvature.

		To do it, we note that the linearized Einstein equations look like
	\bea
	\label{qmn}
	Q_{\mu\nu}&=&\frac{\delta S_{FP}}{\delta h_{\mu\nu}}=-\frac{1}{2}(\pa^{\lambda}\pa_{\mu}h_{\lambda\nu}+\pa^{\lambda}\pa_{\nu}h_{\lambda\mu})+
\frac{1}{2}\eta_{\mu\nu}\pa_{\lambda}\pa_{\rho}h^{\lambda\rho}+\frac{1}{2}\pa_{\mu}\pa_{\nu}h+\frac{1}{2}\Box h_{\mu\nu}-\nonumber\\ &-&
\frac{1}{2}\eta_{\mu\nu}
\Box h=0.
	\eea
	It is evident, and easy to check, that these equations are gauge invariant, $\delta Q_{\mu\nu}=0$, under (\ref{gatra}).
	Hence, we can define the CPT-even gauge invariant action with only fourth derivatives:
	\bea
	\label{four}
	{\cal L}_{four}=\frac{1}{2}b_{\mu}Q^{\mu\nu}b^{\lambda}Q_{\lambda\nu},
	\eea
	where $Q_{\mu\nu}=-(R_{\mu\nu}-\frac{1}{2}Rg_{\mu\nu})$ and this implies
	\bea
	{\cal L}_{four}=\frac{1}{2}b_{\mu}(R^{\mu\nu}-\frac{1}{2}Rg^{\mu\nu})b^{\lambda}(R_{\lambda\nu}-\frac{1}{2}Rg_{\lambda\nu}),
	\eea
where we take only linear terms in $h$ as in eq. (\ref{qmn}).
	
We note that this action differs from that one considered in \cite{HerBel} which involved contraction of Riemann tensors rather than Ricci 
tensors used in our case. A four-derivative Lorentz-breaking term like this was considered in \cite{Ferr2018} in the context of electrodynamics.

Similarly, in the CPT-odd case, we have
	\bea
	\label{odd}
	{\cal L}_{odd}=\frac{1}{2}\epsilon^{\alpha\beta\gamma\delta}b_{\alpha}b^{\mu}Q_{\mu\beta}\pa_{\gamma}b^{\nu}Q_{\nu\delta}.
	\eea
	These modified terms should be added to the usual Lagrangian in eq. (\ref{FP}) of the linearized gravity (\ref{FP}).
	
As in the previous section, our aim consists in search for unusual dispersion relations. Let us calculate first the ingredients of $Q_{\mu\nu}$ 
with use of the decomposition (\ref{decomp}). We have $h=h^{00}-h^{ij}\delta_{ij}=n-2\phi-\chi$, and 
$\pa_{\lambda}\pa_{\rho}h^{\lambda\rho}=-2\nabla^2\dot{n}_L+\nabla^2\chi+\ddot{n}$. Then, we define 
$\pa^{\lambda}\pa_{\mu}h_{\lambda\nu}+\pa^{\lambda}\pa_{\nu}h_{\lambda\mu}\equiv P_{\mu\nu}$. We have:
	\bea
	P_{00}&=&2h_{00}-2\pa_i\dot{h}_{i0}=2(\ddot{n}-\nabla^2\dot{n}_L);\\
	P_{0i}&=&\ddot{h}_{0i}-\pa_j\dot{h}_{ji}+\pa_i\dot{h}_{00}-\pa_i\pa_jh_{j0}=\nonumber\\ &=&\ddot{n}_{iT}+\pa_i\ddot{n}_L-\pa_i\dot{\chi}+\pa_i\dot{n}-\pa_i\nabla^2 n_L;\nonumber\\
	P_{ij}&=&\pa_i\dot{h}_{0j}+\pa_j\dot{h}_{0i}-\pa_i\pa_kh_{kj}-\pa_j\pa_kh_{ki}=\nonumber\\ &=&
	\pa_i\dot{n}_{jT}+\pa_j\dot{n}_{iT}+2\pa_i\pa_jn_L-2\pa_i\pa_j\chi+\nabla^2(\pa_i\xi_{jT}+\pa_j\xi_{iT}).\nonumber
	\eea
	We see that none of this terms involves the physical $\tilde{h}^{ij}$ components, they only enter the $\frac{1}{2}\Box h_{\mu\nu}$ term of $Q_{\mu\nu}$. Hence we see that one has
	\bea
	b_{\mu}Q^{\mu\nu}=\frac{1}{2}b_i\Box \tilde{h}^{ij}\delta^{\nu}_j+(\ldots),
	\eea
	where the dots are for the physically irrelevant components, that is, those other than $\tilde{h}^{ij}$. In this case some of them can 
acquire dynamics but it is common for higher-derivative theories, see e.g. \cite{afraid}. It is important to emphasize that, to get a nontrivial 
impact, the Lorentz-breaking vector $b_{\mu}$ should have essential space-like part which only is contracted to $\tilde{h}^{ij}$.
	As a result, the Lorentz-breaking term (\ref{four}) after integration by parts takes the form
	\bea
	\label{fourfis}
	{\cal L}_{four}=\frac{1}{2}b_ib^k\tilde{h}^{ij}\Box^2 \tilde{h}_{kj}+(\ldots).
	\eea
	So, it remains to study the dispersion relation for the Lagrangian given by the sum of (\ref{fpfis}) and (\ref{fourfis}), which, in the 
relevant sector, yields
	\bea
	{\cal L}_{free}=-\frac{1}{4}\tilde{h}^{ij}\Box \tilde{h}^{ij}+\frac{1}{2}b_ib^k\tilde{h}^{ij}\Box^2 \tilde{h}_{kj}+(\ldots).
	\eea
	The corresponding equation of motion is
	\bea
	-\frac{1}{2}\Box \tilde{h}_{ij}+b_ib^k\Box^2 \tilde{h}_{kj}=(-\frac{1}{2}\Box\delta^k_i+\Box^2b^kb_i)\tilde{h}_{kj}+(\ldots)=0.
	\eea
	As done previously, we can consider that there are only two polarization states. We see that here there are two situations: (i) one has 
simply $\Box \tilde{h}_{ij}=0$ which is the usual Lorentz-invariant situation, with the dispersion relations are the usual ones $E^2=\vec{p}^2$; 
(ii) $(\delta^k_i-2\Box b^kb_i)\tilde{h}_{kj}=0$ which either requires the $b_i$ vector to be directed along the wave propagation direction 
or, in the Fourier representation, requires $\det(\delta^a_i+2p^2b^ab_i)=0$ which enforces $b_a$ to be related with the wave vector $p$ which is 
clearly senseless except of degenerated cases.
	
	It is interesting to compare this situation with the explicitly CPT-breaking case where the quadratic Lagrangian is a sum of the usual 
Lorentz-invariant expression (\ref{FP}) and the CPT-odd term (\ref{odd}), which involves five derivatives. In the same way, we concentrate in 
studying the dynamics of $\tilde{h}_{ij}$. So, we have a sum of the second-order term (\ref{fpfis}) and the fifth-order term
	\bea
	{\cal L}_5=\frac{1}{8}\epsilon_{\alpha\beta\gamma\delta}b^{\alpha}b_i\Box \tilde{h}^{ij}\delta^{\beta}_j\pa^{\gamma}b_k\Box \tilde{h}^{kl}
\delta^{\delta}_l=\frac{1}{8}\epsilon_{\alpha j \gamma l}b^{\alpha}b_i\Box \tilde{h}^{ij}\pa^{\gamma}b_k\Box \tilde{h}^{kl}+(\ldots),
	\eea
	arising from (\ref{odd}). We see that the Lorentz-breaking vector should have a nontrivial space-like part. If it is purely space-like, we 
have after integration by parts
	\bea
	{\cal L}_5=-\frac{1}{8}\epsilon_{mjl}b^mb_ib_k \tilde{h}^{ij}\Box^2 \dot{\tilde{h}}^{kl}+(\ldots),
	\eea
	whose corresponding equation of motion is
	\bea
	-\frac{1}{2}\Box \tilde{h}_{ij}-\frac{1}{4}\epsilon_{mjl}b^mb_ib_k \Box^2 \dot{\tilde{h}}^{kl}+(\ldots)=0.
	\eea
	It is clear that this equation can be rewritten in the form $\Box \Pi_{ij}^{kl}\tilde{h}_{kl}=0$, hence it is compatible with the usual Lorentz-invariant plane wave solutions satisfying the usual equation $\Box 
\tilde{h}_{ij}=0$. As in the previous case, one can have only $b_3\neq 0$ due to the only two polarization states. In this case, the equation 
above will be identically satisfied. So, we see that due to the higher-derivative modes, there is no essential difference between propagation of 
waves in CPT-even and CPT-odd cases.
	
	We can introduce more gauge-invariant terms considering the projection-like operator
	\bea
	\Pi^{\mu\nu}=\eta^{\mu\nu}\Box-\pa^{\mu}\pa^{\nu},
	\eea
	so that $\Pi^{\mu\nu}\Pi_{\nu\lambda}=\Box\Pi^{\mu}_{\alpha}$.
	Then, we consider $\Pi^{\mu\nu}h_{\nu\alpha}$. Its gauge transformation is
	\bea
	\delta\Pi^{\mu\nu}h_{\nu\alpha}=\pa_{\alpha}\Box\xi^{\mu}-\pa_{\alpha}\pa^{\mu}(\partial\cdot\xi).
	\eea
	Afterwards, we construct the vector
	\bea
	K_{\alpha}=b_{\mu}\Pi^{\mu\nu}h_{\nu\alpha},
	\eea
	whose gauge transformation is
	\bea
	\delta K_{\alpha}=\pa_{\alpha}[\Box(b\cdot\xi)-(b\cdot\pa)(\partial\cdot\xi)]=\pa_{\alpha}\Sigma[\xi].
	\eea
	Therefore, the Lagrangian
	\bea
	{\cal L}_{even}=\frac{1}{2}K_{\alpha}\Pi^{\alpha\beta}K_{\beta},
	\eea
	will be gauge invariant since its variation is proportional to $\Pi^{\alpha\beta}\delta K_{\beta}=0$. So, we succeeded to construct the 
higher-derivative aether-like Lorentz-breaking  gauge invariant action for the linearized gravity. 
	
	We note that one can construct a CPT-odd gauge invariant contribution within this prescription as well, it looks like
	\bea
	\label{odd0}
	{\cal L}^{\prime}_{odd}=\epsilon^{\alpha\beta\gamma\delta}b_{\alpha}K_{\beta}\pa_{\gamma}K_{\delta}.
	\eea
	We note that ${\cal L}_{even}$ is of sixth order in derivatives, and ${\cal L}_{odd},{\cal L}^{\prime}_{odd}$ -- of fifth one. Actually, 
${\cal L}_{odd}$ (\ref{odd}) and ${\cal L}^{\prime}_{odd}$ (\ref{odd0}) differ only by irrelevant additive terms which vanish if we put all 
non-physical fields (i.e. all fields other than the transverse-traceless $\tilde{h}_{ij}$) to be equal to zero. In principle, it is possible due 
to the gauge symmetry of these Lagrangians which restricts physical degrees of freedom to $\tilde{h}_{ij}$. Therefore, (\ref{odd0}) and (\ref
{odd}) are physically equivalent.
	We note that these orders in derivatives are very high, corresponding to dimensions-7 and 8 operators (to the best of our knowledge, such 
orders, except of essentially nonlocal models, were studied only within of a very specific context of Rashba coupling \cite{Rashba, Nasc}; we 
note that in \cite{KosLi}, the table includes only operators with dimensions up to 6), and, moreover, they cannot be decreased without 
introduction of undesired nonlocal terms  involving negative degrees of $\Box$, which are rather dangerous from the unitarity/causality viewpoint. 
	Besides, the corresponding full-fledged contributions to the action expressed in terms of the Riemann curvature tensor and its covariant 
derivatives are not known, and search for them is a nontrivial problem.

\section{Summary}

We discussed the modifications of dispersion relations for various LV extensions of gravity and corresponding changes in the plane wave 
solutions. We demonstrated explicitly that only in certain cases the dispersion relations turn out to be essentially different from Lorentz 
invariant ones. We showed that the dispersion relations continue to be the usual ones for a specific class of Lorentz-breaking extensions of the gravity, namely, the aether-like ones characterized by one constant vector. Clearly, it establishes the question about a form of dispersion relations in more involved cases. Certainly, gauge-breaking LV extensions of gravity, discussed in \cite{KosGB}, require more detailed studies. In particular, it is interesting to construct more involved LV extensions of gravity which could display unusual dispersion relations whose possibility has been demonstrated in \cite{Mewes}. 

Another result of our study is the formulation of prescription allowing for generating gauge invariant LV extensions of the Einstein-Hilbert
lagrangian for the weak field with any arbitrary number of derivatives. We expect that such extensions can be useful for studying certain 
physical phenomena.
 
A possible extension of this paper could consist in studies of plane wave solution on a nontrivial curved background. Another possible 
development of this study can consist in detailed consideration of massive LV gravity, while up to now most studies of massive gravity were 
devoted to Lorentz-invariant case, see \cite{Hint} and references therein. Besides, it is natural to expect that nontrivial 
phenomena taking place within wave propagation discussed in the paper can be used in future gravitational wave observations in order to find LV 
extensions of gravity which could be more appropriate from the experimental viewpoint.

\vspace{6pt} 




{\bf Acknowledgments.} This work was partially supported by Conselho
Nacional de Desenvolvimento Cient\'{\i}fico e Tecnol\'{o}gico (CNPq). The work by A. Yu. P. has been supported by the
CNPq project No. 301562-2019/9.

\end{document}